\documentclass[letterpaper,twocolumn,10pt]{article}
\usepackage{usenix-2020-09}

\usepackage{graphicx} 
\usepackage{xspace}
\usepackage{enumitem}
\usepackage{subfig}
\usepackage{subcaption}
\usepackage{booktabs}
\usepackage{xcolor}
\usepackage{pifont}
\usepackage{tikz}
\usetikzlibrary{arrows.meta, automata, positioning, shapes.geometric}
\usepackage{tabularx}
\usepackage{array}
\usepackage{comment}
\usepackage{algorithm}
\usepackage{algpseudocode}
\usepackage{amsmath}
\usepackage[capitalise,nameinlink,noabbrev]{cleveref}

\crefname{section}{\S}{\S{}}
\crefname{figure}{Fig.}{Fig.}
\crefformat{section}{\S#2#1#3}

\usepackage[small,compact]{titlesec} 

\newcommand{\tinyskip}{\vspace{3pt}}

\newcommand{\mypar}[1]{\tinyskip\noindent\textbf{#1.}\xspace}

\newcommand{\myparij}[1]{\noindent\textbf{\small#1}\xspace}

\begin{document}

\newcommand{\sysname}{DPC\xspace}
\newcommand{\todo}[1]{{\color{red}\small[todo: #1]}}
\newcommand{\YZ}[1]{{\color{purple}\small[yz: #1]}}
\newcommand{\SB}[1]{{\color{blue}\small[sb: #1]}}
\newcommand{\GN}[1]{{\color{olive}\small[gn: #1]}}
\newcommand{\etal}{et al.\xspace}
\newcommand{\JE}[1]{{\color{cyan}\small[je: #1]}}
\newcommand{\AT}[1]{{\color{orange}\small[AT: #1]}}

\newcommand*\circled[1]{\tikz[baseline=(myanchor.base)] \node[circle,fill=.,inner sep=1pt] (myanchor) {\color{-.}\bfseries\footnotesize #1};}
\newcommand*\squared[1]{%
  \tikz[baseline=(myanchor.base)]%
    \node[
      rectangle,
      fill=red,
      text=white,
      minimum size=1.0em,
      inner sep=0pt
    ] (myanchor)
    {\bfseries\footnotesize #1};%
}
\newcommand*\diamonded[1]{%
  \tikz[baseline=(myanchor.base)]%
    \node[
      diamond,
      fill=green!50!black,,
      text=white,
      inner sep=1pt,
      minimum size=1.0em
    ] (myanchor)
    {\bfseries\footnotesize #1};%
}

\newcommand*\circledw[1]{\tikz[baseline=(myanchor.base)] \node[circle,fill=white,inner sep=0.9pt,draw=black,line width=0.2mm] (myanchor) {\color{black}\bfseries\footnotesize #1};}

\title{DPC: A Distributed Page Cache over CXL}

\author{
{\rm Shai Bergman}\\
Huawei Technologies Switzerland AG\\
shai.aviram.bergman@huawei.com
\and
{\rm Zhe Yang}\thanks{Corresponding author.}\\
Huawei Technologies Co., Ltd\\
yangzhe.ac@outlook.com
\and
{\rm Julien Eudine}\\
Huawei Technologies Switzerland AG\\
julien.vincent.eudine@huawei.com
\and
{\rm Giorgio Negro}\\
Huawei Technologies Switzerland AG\\
giorgionegro@proton.me
\and
{\rm Onur Mutlu}\\
ETH Zurich\\
omutlu@gmail.com
\and
{\rm Arash Tavakkol}\\
Huawei Technologies Switzerland AG\\
arash82ir@gmail.com
\and
{\rm Ji Zhang}\footnotemark[\value{footnote}]\\
Huawei Technologies Co., Ltd\\
dr.jizhang@huawei.com
}

\date{}
\maketitle
\hypertarget{Hfootnote.1}{} 

\begin{abstract}

Modern distributed file systems rely on uncoordinated, per-node page caches that replicate hot data locally across the cluster. While ensuring fast local access, this architecture underutilizes aggregate cluster DRAM capacity through massive data redundancy and incurs prohibitive coherence overhead via heavyweight, lock-based protocols. In this paper, we focus on the design of a distributed page cache that treats the entire cluster's main memory as a single cache budget while preserving standard file-system interfaces and semantics. We present Distributed Page Cache (\sysname), an OS-level, distributed page cache built on top of Compute Express Link (CXL) 3.0 memory semantics. \sysname enforces a single-copy invariant at page granularity: each file page has exactly one owner node holding the sole resident DRAM copy, and other nodes access it via CXL-based remote mappings rather than creating replicas of the page. \sysname is implemented end-to-end on a CXL-based emulation framework that models multi-host CXL 3.0 memory fabrics, enabling detailed evaluation in the absence of widespread hardware. Across real-world and representative data-sharing workloads, \sysname delivers speedups of up to 12.4$\times$, with a geometric-mean speedup of 5.6$\times$.

\end{abstract}
\section{Introduction}

The \emph{page cache} is a fundamental component of modern OSs for bridging the orders-of-magnitude latency gap between CPU and persistent storage. In most modern OSs, an in-memory cache of file and block data exploits temporal and spatial locality of reference in typical workloads, allowing the system to mask much of the latency of slower storage devices~\cite{ding2007buffer, jiang2005dulo, pai2000io, smith1985disk}. By caching hot blocks in main memory, the OS kernel satisfies read requests via low-latency memory hits and amortizes write costs through write-back buffering. Although the storage landscape has evolved from rotational media to high-performance NVMe SSDs, the latency gap between main memory (tens of nanoseconds) and persistent storage (tens to hundreds of microseconds) persists by orders of magnitude. Consequently, the page cache remains a critical component in the I/O datapath, dynamically scaling with available system main memory to maximize the hit rate. Modern OSs aggressively allocate surplus DRAM to the page cache, employing sophisticated reclamation, clustering, and read-ahead policies to mask storage latency~\cite{kroeger1996predicting,jacob2010memory,kroeger1999case}. Recognizing this efficiency, high-performance data platforms, such as Kafka~\cite{kreps2011kafka}, Cassandra~\cite{lakshman2010cassandra}, and RocksDB~\cite{dong2021rocksdb}, are architected to leverage the OS page cache, explicitly eliminating custom application-level caching~\cite{kreps2011kafka,dai2024symbiosis,Dong2021FAST,carpenter2022cassandra}. As a result, the page cache has become a first-class performance substrate, not just a best-effort optimization~\cite{sosp25-cache-ext,lee2023p2cache,peng2024scalacache}. 

In distributed clusters and High-Performance Computing (HPC) file systems, however, this abstraction is almost always instantiated as independent, per-node caches layered on top of a shared parallel file system such as GPFS/Spectrum Scale~\cite{schmuck2002gpfs} or Lustre~\cite{braam2019lustre}. Each client runs an unmodified kernel that maintains its own page cache; the parallel file system enforces consistency using distributed lock managers or token systems, but it does not provide a true cluster-wide cache. For example, GPFS uses a token-based lock manager to control which clients may cache which file ranges, and tokens must be revoked before another client can read or write that range~\cite{schmuck2002gpfs}. Lustre similarly maintains client-side data and metadata caches but relies on the Lustre Distributed Lock Manager (LDLM) and metadata write-back cache to keep those caches consistent~\cite{schwan2003lustre}. 
In this model, each client’s page cache is an island: nodes have no visibility into each other’s cached contents, and \emph{global} caching emerges only as the accidental union of many disjoint local caches; even when a system attempts to pool caches at a higher layer, CPUs still cannot directly access each others' main memory and must transfer entire pages over the network. 

Decades of work on \emph{cooperative} and \emph{global} caching underscore both the necessity and the difficulty of treating client memory as a shared resource. The cooperative caching technique of Dahlin et al.~\cite{cooperative-osdi-94} showed that coordinating client caches and allowing cache misses to be satisfied from remote client memory can halve the number of disk accesses and improve read latency by up to 73\%. Subsequent work generalized this idea into locality-aware cooperative cache managers that form a unified level of cache above the server, again showing substantial reductions in disk traffic and improved response times on real traces~\cite{jiang2006locality}. More recently, SP-Cache revisits the caching problem in shared, cluster-wide in-memory caches for data-intensive analytics, rather than OS-level page caches~\cite{spcache-sc-18}. It shows that naive replication or erasure-coded caching of popular files wastes memory, and instead selectively partitions hot files and distributes disjoint partitions across cache servers. This redundancy-free placement reduces mean and tail access latency by up to 40\% while using about 40\% less memory, reinforcing the case for coordinated, cluster-wide caching rather than independent local caches.

\begin{figure}
\captionsetup{aboveskip=4pt, belowskip=0pt}
    \centering
    \includegraphics[width=1.0
    \linewidth]{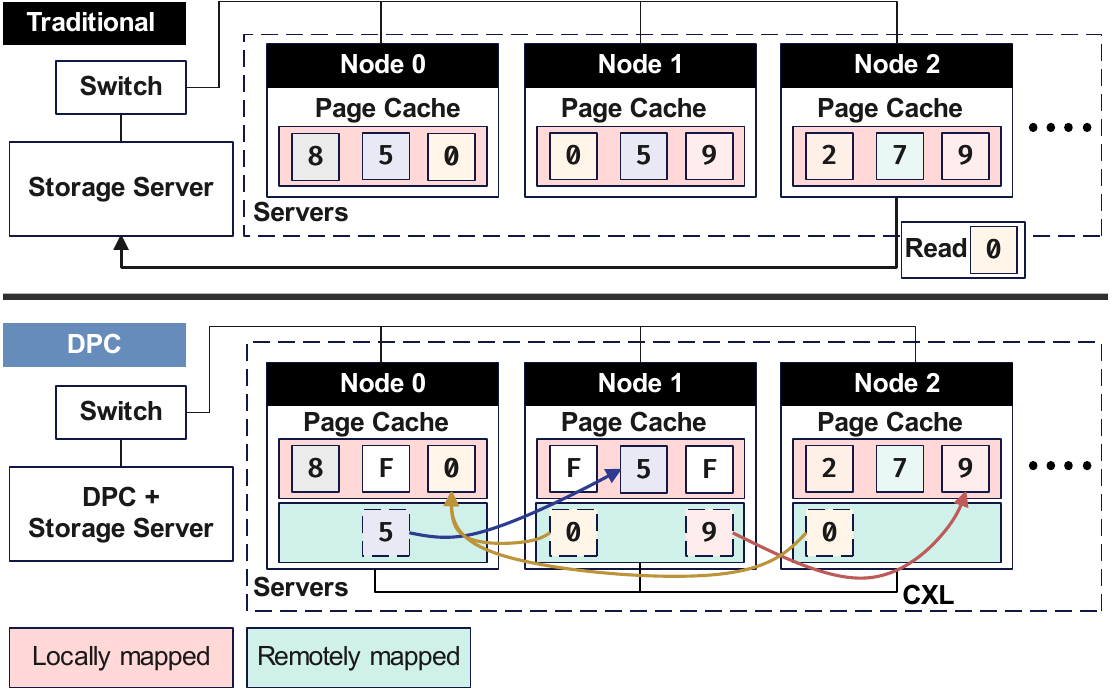}
    \caption{Baseline architectures. Top: traditional cluster where each node maintains an independent page cache, replicating hot pages across DRAM. Bottom: A DPC-enabled cluster where nodes share a page cache; pages may be mapped locally or remotely ("F" marks frames freed by remote mappings, increasing effective cluster-wide cache capacity).}
    \label{fig:systemcontext}
\end{figure}

Despite this evidence, today's distributed and HPC file systems largely remain in the per-node page caching regime, which introduces \textbf{two fundamental problems}.\\
First, independently managed client caches (see \cref{fig:systemcontext}) waste memory on redundant replicas of hot data. Popular files are accessed by many nodes, and because each client makes caching decisions in isolation, the same hot blocks are cached on tens of nodes. From a cluster-wide perspective, this replication does not materially increase the hit rate once a few copies exist; the marginal benefit of the N-th replica is small, but its memory cost is linear in N. In other words, the aggregate DRAM budget in a cluster is not being used to capture global locality; it is being spent on storing identical local copies.\\
Second, per-node caches make cache coherence and consistency control increasingly costly as clusters scale to larger sizes. Parallel file systems must track which clients might hold cached copies and must invalidate or revoke access on writes or conflicting metadata updates. Treating each client cache as an island either invites stale reads or forces conservative policies, such as write-through, frequent invalidations, or even cache bypass, that erode the latency benefits of caching. Attempts to enforce stronger consistency, like distributed locking and lease or callback schemes, introduce their own overheads. Experience with production-scale systems shows that the control-plane work to keep client-side metadata caches consistent can become overwhelming as clusters grow, creating bursts of renewals or invalidations that can overload the servers~\cite{infinifs-fast-22}. The common fallback, i.e., disabling caching for hot metadata or using fixed-block, synchronous data writes via the \texttt{O\_DIRECT} flag, avoids coherence storms but introduces repeated lookups and high-latency IO operations~\cite{assise-osdi-20}. Consequently, without a coherent, efficient, cluster-wide cache abstraction, systems face a lose-lose trade-off: either accept stale data risk and complex failure modes, or give up a substantial fraction of the potential page cache hit rate. A distributed cache manager that provides scalable coherence semantics is therefore essential to unlock low-latency reads and high hit rates without the prohibitive coordination costs seen in uncoordinated designs. 

Taken together, these observations motivate a distributed page cache at the OS level that (1) uses the cluster's memory as a single  cache, minimizing redundant replicas, and (2) provides scalable coherence semantics without incurring prohibitive software coordination overhead. Existing cooperative caching systems and cluster caches (e.g.,~\cite{cooperative-osdi-94,jiang2006locality,spcache-sc-18,indexfs}) demonstrate the potential, but they sit at the file-system or library layer, rely on software RPCs over commodity networks for every remote hit, and are typically optimized for specific workloads or file systems rather than being a general solution like the OS kernel page cache. 
To our knowledge, no prior work provides a transparent, OS-integrated distributed page cache that both avoids redundant copies and offers scalable coherence while preserving conventional file system semantics.

We propose \textbf{Distributed Page Cache (\sysname)} to address these limitations by aggregating memory across machines into a single logical page cache pool and enforcing a non-replicated placement policy at page granularity. As shown in \cref{fig:systemcontext}, in \sysname, each cached page has a unique owner node responsible for holding the canonical in-memory copy. A node that misses in its local page cache first consults a lightweight directory (logically similar to a cache tag store) to determine whether any peer holds the requested page. If a cached copy exists, \sysname routes the request directly to the owning node and serves the read via low-latency remote memory access, avoiding disk or remote file-server I/O entirely. Only if the directory reports a miss does the node access the underlying storage, install the page into the distributed cache, and potentially evict some other page according to its local replacement policy designed to give each node full control over its local memory management. By construction, the cluster's aggregate DRAM is used to store disjoint subsets of the working set rather than redundant replicas, and read-sharing across nodes is satisfied at remote-memory latency rather than storage latency. \sysname thus exposes to applications the familiar semantics of a local page cache while internally operating as a distributed cache.

The \textbf{key enabler} for \sysname's design is \textit{Compute Express Link} (CXL) 3.0~\cite{cxl-3.1-spec}, which extends earlier point-to-point CXL links into a hardware-coherent memory fabric spanning multiple hosts. CXL 2.0 introduced pooled and fabric-attached memory across hosts~\cite{cxl-2.0-spec}, while CXL 3.0 adds multi-level switching and, critically, hardware-managed coherence for shared address ranges. In CXL, physically distinct systems can issue load/store operations to common memory regions, and the fabric maintains coherence at cache-line granularity, e.g., via invalidations and snoop filters, similar to a NUMA system. This coherence is \emph{memory-centric}: it governs CPU caches over a shared physical address space, not higher-level file-system pages. By contrast, a distributed page cache for a cluster built on top of CXL operates at the file/page abstraction and must still manage placement, eviction, and consistency of cached file blocks across nodes. \sysname leverages CXL 3.0 by mapping the node's exported page cache pages into a shared CXL address space, so that a \emph{remote cache hit} is satisfied via a CXL memory transaction rather than a software-mediated RPC. The CXL fabric, not the file system, enforces cache-line coherence within these regions, while \sysname's distributed page cache protocol remains responsible for higher-level page ownership and visibility. In effect, CXL 3.0 supplies a distributed, coherent hardware substrate, and \sysname builds on it to transform redundant per-node page caches into a logically single, cluster-wide page cache operating at remote-memory speeds.

To summarize, the contributions of this work are as follows:
\begin{enumerate}[noitemsep, partopsep=0pt,topsep=0pt,leftmargin=0pt,itemindent=12pt]
    \item We present the design of \sysname, a transparent, distributed page cache that aggregates memory across nodes over a CXL fabric. \sysname preserves existing application and file-system interfaces (e.g., POSIX I/O and mmap) while supporting optional strong semantics, including POSIX-like coherence for file data and shared-memory-style consistency for mapped regions, without requiring application changes.
    \item We design a page-granular cache directory and a software-defined coherence protocol for file-backed pages that spans multiple nodes. This design enables low-latency remote cache hits and coordinated eviction/invalidation at page granularity.
    \item We implement \sysname end-to-end on a CXL-based emulation framework that models CXL 3.0-style memory fabrics, enabling detailed experimentation and validation of our design in the absence of widely available multi-host CXL 3.0 hardware.
    \item We evaluate \sysname using real-world and representative data-sharing workloads, comparing it against state-of-the-art distributed file-access mechanisms, including VirtioFS~\cite{hajnoczi2019virtiofs,russell2008virtio}, NFS~\cite{sandberg1985nfs}, and JuiceFS~\cite{juicedata2021juicefs}. Across these workloads within a multi-node setup, \sysname achieves speedups of up to 12.4$\times$, with a cross-workload geomean speedup of 5.6$\times$.
\end{enumerate}

\section{Distributed Memory Access Over CXL}  

\subsection{From explicit RDMA to coherent CXL}
Remote memory has long been attractive as a way to build cluster-wide caches: if a cache miss can be served from a peer's DRAM instead of local storage, the latency reduction is substantial. \textit{Remote Direct Memory Access} (RDMA) over InfiniBand~\cite{dragojevic2014farm} or RoCE~\cite{guo2016rdma} made such designs feasible by exposing low-latency remote access. Therefore, prior work on cooperative caching and distributed shared memory has relied on RDMA as the primary substrate~\cite{mitchell2013using,dragojevic2014farm,nsdi17-infiniswap}. However, RDMA exports non-coherent, non-byte-addressable memory: remote regions are visible only through explicit RDMA verbs, not as part of the CPU's load/store address space. The OS thus treats remote buffers as opaque endpoints that must be managed explicitly in software, and every coherence action, such as tracking ownership, invalidating copies, or flushing dirty data, must be implemented by custom protocols layered atop message-passing. RDMA is therefore well-suited for bulk, explicit dta movement, but ill-suited as the foundation for a transparent page cache or true memory sharing/extension that should behave like local physical memory to the kernel.


CXL alters this landscape by offering a cache-coherent interconnect. CXL defines three protocols operating simultaneously on a single link: \texttt{CXL.io} for discovery/configuration; \texttt{CXL.cache} for device caching of host memory; and \texttt{CXL.mem}, which enables the host processor to map device memory into its physical address space using standard load/store semantics. CXL 3.0~\cite{cxl-3.0-spec} generalizes prior CXL designs into a true memory fabric and, critically, introduces hardware-managed coherence across multiple hosts, allowing physically distinct systems (i.e., separate OS instances) to concurrently access shared memory regions without software intervention. These capabilities are pivotal for distributed page caching. First, CXL memory appears to the OS as byte-addressable DRAM rather than a network endpoint; its access latency resembles more of a NUMA node rather than a block device. Second, CXL 3.0's shared-memory semantics allow multiple nodes to map the same physical pages while relying on hardware to maintain cache coherence. This combination positions CXL as a natural substrate for exposing remote cache pages as if they were local. In the remainder of this paper, we use the term CXL to refer specifically to the \texttt{CXL.mem} protocol and the coherence semantics of the CXL 3.0 specification.  

\subsection{Key challenges}
\label{section:challenges}
Implementing a distributed page cache over CXL faces fundamental challenges at the boundary between hardware and OS requirements. \\
\textbf{\small 1. The semantic and granularity gap.} There is a fundamental impedance mismatch between the fine-grained, memory-centric coherence of CXL and the coarse-grained, file-centric consistency required by the OS page cache. While CXL strictly enforces coherence at the granularity of cache lines (e.g., 64~B) to synchronize CPU caches with device memory, the page cache manages data in 4~KB units (or larger in the future) 
governed by higher-level file system semantics such as POSIX file semantics and consistency. 
Crucially, the hardware is oblivious to these software abstractions: the CXL fabric perceives only physical addresses and coherence states, remaining blind to which collection of cache lines constitutes a logical page, which backing file those lines belong to, or whether a specific frame contains valid file data versus stale remnants from a prior allocation. To the fabric, shared memory is simply a range of anonymous bytes; the binding between these physical lines and the logical file system, page cache page locations, and OS paging must be synthesized entirely by the OS.\\ 
\textbf{\small 2. The reclamation-coherence tension.} A distributed page cache must reconcile local memory autonomy with global coherence. In a CXL-based cluster, a physical page owned by one node may be mapped by multiple remote peers, and some of those mappings may be writeable. At the same time, each node experiences its own memory pressure and must be able to reclaim page cache page according to its own policies, without blocking on cluster-wide negotiation. This creates a fundamental tension: from one node's perspective, a page may appear reclaimable (no local references, low utility), yet that same physical page may still be actively mapped or dirty on one or more remote nodes. Reclaiming it unilaterally risks dangling mappings and silent data loss; refusing to reclaim it until every potential remote user has explicitly released it risks turning remotely mapped pages into effectively pinned memory that cannot be freed when local pressure is high. What is needed instead is a form of \emph{deterministic reclamation}: once a node decides that a page should be reclaimed, there must be a clear, bounded sequence of steps that leads to that page becoming free in its local memory, while still preserving correctness for any outstanding remote mappings. \\
\textbf{\small 3. Stringent latency margins and control-plane overhead.} Even if semantic and reclamation issues are resolved, the system must justify its existence within tight latency budgets. The primary motivation for a distributed page cache is to bridge the gap between local DRAM and backend storage; however, modern NVMe SSDs have shrunk this gap significantly. Consequently, the latency budget for control-plane operations (directory lookups, ownership transfers, invalidations) is extremely thin. Any mechanism that requires multiple software round-trips can easily negate the benefits of remote caching. This challenge is compounded by interconnect contention. In real systems, coherence traffic often competes with bulk data transfers for link bandwidth. If invalidation messages are subjected to head-of-line blocking behind large page migrations, nodes may observe stale data or suffer stalled reads. Thus, the protocol must be lightweight enough to operate at network/memory timescales, not storage timescales, ensuring that the cost of coherence does not erode the gains of the cache hit. 

\newcolumntype{L}[1]{>{\raggedright\arraybackslash}p{#1}}
\newcolumntype{C}[1]{>{\centering\arraybackslash}p{#1}}
\newcolumntype{X}{>{\raggedright\arraybackslash}X}
\newcommand{\xmark}{{\color{red}\ding{55}}}
\newcommand{\cmark}{{\color{green!70!black}\ding{52}}}


\section{DPC Design}
\label{sec:design}








DPC comprises two cooperating components: a \textbf{DPC Directory}, and a \textbf{DPC Client}. The directory maintains page-granular state for the distributed cache and coordinates ownership, invalidations, and reclamation across nodes. The client integrates with the local file system, memory management, and page cache, mapping remote pages over CXL while participating in the directory protocol on accesses and evictions. Together, these components let DPC expose the familiar abstraction of a local page cache while operating over a shared, fabric-attached cache.

\subsection{DPC Cache directory design}
At the heart of our design is a cache directory that sits \textit{logically} on the storage server (shown in \cref{fig:systemcontext}), mediating all file-system requests from attached nodes and providing the missing control-plane for a distributed page cache over CXL. The directory is responsible for addressing the two core challenges identified in section~\ref{section:challenges}: (1) the semantic and granularity gap between CXL's cache-line coherence and the kernel's page-level page cache, and (2) the tension between local memory reclamation and cross-node coherence. To bridge the first gap, the directory maintains page-granular metadata for every cached file page across nodes and enforces a \textbf{single-copy} invariant: at any point in time, exactly one node is the owner of a cached page, and no other in-memory replicas of that page exist. \cref{fig:systemcontext} illustrates this single-copy invariant. As long as ownership does not change, all other nodes treat that page as remotely owned and, after consulting the directory, must direct their reads and writes to the owner node rather than caching their own copies.\\
Directory operations are atomic at the page level, ensuring well-defined, consistent transitions in ownership and mapping state while allowing concurrent updates to different pages for high throughput; the CXL fabric then provides fine-grained cache-line coherence among nodes that map the owner's page. To address the reclamation-coherence tension, the directory participates in every eviction: before a node can reclaim a physical page, the directory ensures that all dependent remote mappings have been torn down and that ownership and dirty state are resolved, preventing dangling references while allowing each node to make independent reclamation decisions. 

\subsubsection{Page-level directory state machine}
\label{subsec:directory}
The cache directory maintains, for each cached logical file page, a state machine that describes how that page is accessed across the cluster. Each directory entry of an actively cached page stores a vector of \textbf{per-node states} whose combination defines the cluster-wide view of that page. Concretely, the entry records (i) which node, if any, currently owns the sole resident DRAM copy of the page, and (ii) which other nodes, if any, hold active mappings to that page. 

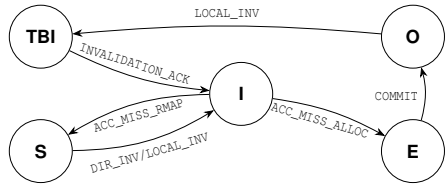
\begin{figure}
\captionsetup{aboveskip=4pt, belowskip=0pt}
\vspace{-1.0\baselineskip} 
\centering
\resizebox{0.7\linewidth}{!}{
\begin{tikzpicture}[%
  >=Stealth,
  every state/.style={draw,circle,minimum size=11mm,
  font=\sffamily\bfseries},
  every edge/.style={draw=black,thick,-{Stealth}}]

  \node[state] (I)  at (1,1.0) {I};  
  \node[state] (O) at (4,2.0) {O}; 
  \node[state] (S)  at (-2.5,0) {S};  
  \node[state] (E)  at (4,0) {E};  
 \node[state] (TBI) at (-2.5,2.0) {TBI};  

  \draw[->,bend right=15]
(I) to node[pos=0.5,below,sloped]
   {\scriptsize\texttt{ACC\_MISS\_RMAP}}
(S);

  \draw[->,bend right=15]
(S) to node[pos=0.5,below, sloped]
   {\scriptsize\texttt{DIR\_INV/LOCAL\_INV}}
(I);

  \draw[->,bend left=10]
(I) to node[pos=0.43,below,sloped]
   {\scriptsize\texttt{ACC\_MISS\_ALLOC}}
(E);

  \draw[->,bend right=10]
(TBI) to node[pos=0.45,above,sloped]
   {\scriptsize\texttt{INVALIDATION\_ACK}}
(I);

  \draw[->,bend right=15]
(E) to node[pos=0.45,left] {\scriptsize\texttt{COMMIT}}
(O);

  \draw[->,bend right=5]
(O) to node[pos=0.5,above,sloped]
   {\scriptsize\texttt{LOCAL\_INV}}
(TBI);

\end{tikzpicture}
}
\caption{Page-level directory state machine. For each page, a node's state is one of five values: Invalid (I), transient Exclusive (E), Owner (O), Shared (S), or To-Be-Invalidated (TBI).}
\label{fig:fsm}
\end{figure}

As shown in \cref{fig:fsm}, we distinguish five per-node states:\\
\textbf{\small Invalid (\textsf{I}).} The node has no local resident copy of the page and no active remote mapping to it; from this node's perspective, the page is absent. \textsf{I} is the default state for every node before the page is created or accessed.\\ 
\textbf{\small Exclusive (\textsf{E}).} The \textsf{E} state is a transient reservation state, used whenever there is currently no owner for a page in the cluster (all nodes are in \textsf{I}) and some node is about to become the new owner, either by loading the page from storage or by performing a full-page write (including creation or complete overwrite). When the directory grants \textsf{E} to a node, it does not assume that a valid resident copy already exists; instead, it records that this node has the exclusive right to install the next resident copy. Two properties follow. First, while a page is in \textsf{E}, no valid cached copy exists anywhere in the cluster, and the requester or the file system is responsible for materializing the page contents: the requester if it is performing a full-page write, or the file system if the page must be fetched from the backing store. Second, the \textsf{E} state prevents all other nodes from acquiring a mapping or reading any prior contents. Because the requester does not yet have valid data in its DRAM, it would be unsafe to let other nodes access the page while installation is in progress. Only after the new contents have been installed does the directory atomically promote the page to Owner (\textsf{O}) on that node.\\
\textbf{\small Owner (\textsf{O}).} Exactly one node is the owner of the page and holds the only resident copy in its local DRAM. This node is responsible for the page's lifetime and writeback to the backing store when eviction is required.\\
\textbf{\small Shared (\textsf{S}).} The \textsf{S} state describes remote mappings of an owned page. A node in \textsf{S} does not hold its own DRAM copy; rather, its CPUs access the owner's DRAM page over the CXL fabric using a read-write mapping. From the directory's perspective, the page remains owned by a single node in \textsf{O}, while any number of other nodes may concurrently be in \textsf{S}, each mapping the same physical page exported by the owner.\\
\textbf{\small To Be Invalidated (\textsf{TBI}).} The \textsf{TBI} state represents a page that is in the process of being torn down. When the current owner intends to evict a page, it must first ensure that no remote mappings remain and that any dirty data is safely reflected in the backing store. The directory therefore enters \textsf{TBI} for that node while invalidation is in progress: it marks that this page is no longer available for new mappings, initiates invalidation messages to all nodes that were in \textsf{S}, and waits for acknowledgments to complete. Only after all sharers have dropped their mappings and the owner has resolved dirty state does the directory allow the page to transition to \textsf{I} (no cached copy). \textsf{TBI} is necessary to avoid races in which some nodes continue to use a page even though it has been reclaimed or reassigned.

As shown in \cref{fig:fsm}, transitions between these states are driven by a small set of directory events, each corresponding to a high-level operation on a page:\\
\textbf{\small 1. \texttt{ACC\_MISS\_ALLOC}.} Triggered when a node accesses a page and the directory finds no resident copy anywhere in the cluster (all nodes are in \textsf{I}). \\
\textbf{\small 2. \texttt{COMMIT}.} Triggered when the page initialization is complete on the node in the \textsf{E} state.\\ 
\textbf{\small 3. \texttt{ACC\_MISS\_RMAP}.} Triggered when a node accesses a page that is owned by exactly one other node (\textsf{O}).\\
\textbf{\small 4. \texttt{LOCAL\_INV}.} Triggered when the owner evicts the page or a sharer invalidates its remote mapping.\\ 
\textbf{\small 5. \texttt{DIR\_INV}.} Triggered when an \textsf{S}-state node acknowledges invalidation, indicating it no longer has access to the page.\\ 
\textbf{\small 6. \texttt{INVALIDATION\_ACK}.} Triggered after the directory has received acknowledgments from all \textsf{S}-state nodes, notified the owner, and the owner has completed any required write-back.

\subsubsection{Functional components of the directory}
\begin{figure}
\captionsetup{aboveskip=4pt, belowskip=0pt}
\vspace{-0.5\baselineskip} 
\centering
\includegraphics[width=1\linewidth]{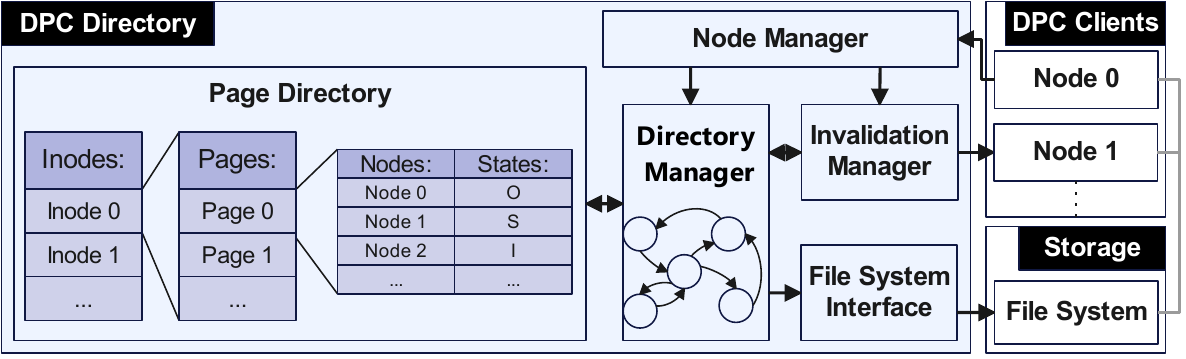}
\caption{The components of the DPC Directory.}
\label{fig:directory}
\end{figure}

\cref{fig:directory}
illustrates the main components of the directory:\\
\textbf{\small Page Directory} organizes metadata as a two-level hash map
keyed by file inode and page index. Each entry stores the per-node
state vector for a single, cached, logical page as defined in~\cref{subsec:directory},
recording the current owner, the set of sharers, and the owner’s
page-frame number. Directory operations touch only the entry for
the accessed page, so 
concurrent requests for different pages can scale.\\
\textbf{\small Directory Manager} implements the page-level protocol
and maintains the single-copy invariant. Conceptually, the
Directory Manager exposes two logical operations to clients:
(i) lookup-and-install for data misses, which
either allocates a fresh entry via a transient state and assigns ownership, or returns the
existing owner and page-frame number; and (ii) reclaim and
invalidation, which coordinate the invalidation of remote mappings
when an owner evicts a page.\\
\textbf{\small Communication Components} connect the directory to the rest of the
system. The \emph{Node Manager} tracks compute nodes, multiplexing
per-node connections and attaching node identifiers. The \emph{Invalidation Manager} orchestrates
owner-initiated invalidations, batching
requests per page and tracking acknowledgments from sharers.
Finally, the \emph{File System Interface} forwards I/O requests to the backing
store on directory misses. Together these components provide a centralized
control plane that presents a per-page interface to DPC clients.

\subsection{DPC Client design}

\begin{figure}
\captionsetup{aboveskip=4pt, belowskip=0pt}
\vspace{-0.5\baselineskip} 
\centering
\includegraphics[width=0.89\linewidth]{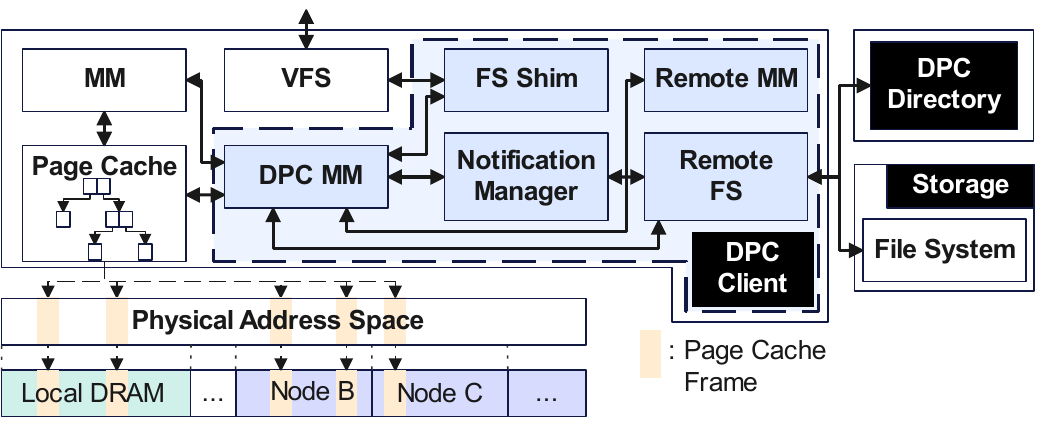}
\caption{The components of the DPC Client.}
\label{fig:client}
\end{figure}

\cref{fig:client} illustrates the DPC client, which runs on each compute node and bridges the local kernel with the CXL fabric. The client integrates transparently with existing file and memory management by interfacing with the virtual file system (VFS), the kernel layer that mediates POSIX file-system calls. Its components are organized around three responsibilities: (i) presenting a standard file-system interface to the VFS, (ii) coordinating with the DPC directory for placement, lookup, and both local and remote page reclamation, and (iii) mapping remote cache pages over CXL into the local physical address space and inserting them into the page cache. Our design makes a remote page appear to the kernel as an ordinary page-cache page while its physical frame resides on a peer and is accessed via coherent CXL loads and stores instead of software RPCs. Each component is explained below:\\
\textbf{\small FS Shim} interposes on file-system operations for DPC-capable mounts and forwards page-cache misses and eviction events to the DPC client logic.\\
\textbf{\small DPC MM (Memory Manager)} is the kernel-side component that connects the local page cache to the rest of the system. DPC MM tracks which cached file pages are enrolled in DPC, issues directory lookups on misses, updates page-cache metadata when ownership or mappings change,
and coordinates the local eviction of owned pages through Linux's generic memory management subsystem (MM) with the directory.\\
\textbf{\small Notification Manager} is the key to giving each \textit{remote node} independent control over its memory while preserving correctness. It delivers asynchronous invalidation events from the directory to the DPC MM, which promptly unmaps affected pages from local page tables and evicts them from the page cache. Thus, whenever a peer decides to reclaim a physical page, that decision is followed by a deterministic sequence of directory-initiated invalidations before the page is freed.\\
\textbf{\small Remote FS} is the client’s control-plane endpoint to the cache directory and the backing storage.\\ 
\textbf{\small Remote MM (Memory Manager)} exposes DRAM exported by remote DPC nodes as reserved regions in the local physical address space. As shown in the lower half of \cref{fig:client}, a node’s physical address space consists of local DRAM plus one or more CXL-mapped ranges from peers. Remote MM enumerates these ranges and converts directory responses (owner and remote address) into local physical frame numbers, which the DPC MM then installs in the page cache as if they were backed by local DRAM.

\section{Implementation}
\label{sec:implementation}
DPC is implemented on top of \textbf{Virtiofs}~\cite{virtiofs-project-2025}, but the design is not tied to that particular file system. Virtiofs is a shared file system for Linux that lets a client mount a directory tree exported by a server-side daemon, using a FUSE-like request/response protocol~\cite{vangoor2017fuse}. This makes it an attractive substrate for DPC: it is a common, standardized platform, already integrated with the VFS layer, and naturally supports offloading file operations to user-space services and accelerators~\cite{DPFS}. Communication between client and server in Virtiofs uses \textbf{virtqueues} (ring buffers) for request and completion messages; DPC builds on this by provisioning dedicated virtqueues for different control paths (e.g., directory operations and invalidations). On the client, the in-kernel Virtiofs driver translates VFS operations into FUSE requests and forwards them to the \texttt{virtiofsd} daemon, which executes the corresponding operations on the backing file system. Modern SmartNICs/DPUs can emulate Virtiofs PCIe devices and expose them directly to attached hosts~\cite{nvidia_doca_vfs_2024,DPFS}, allowing the same protocol to run unchanged in offloaded  deployments.



The goal of our design is to address the efficiency challenges outlined in \cref{section:challenges} while meeting the functional requirements of a transparent, OS-level distributed page cache. We implement a DPC prototype on Linux 6.12.5 by extending the upstream in-kernel Virtiofs file system and the standard user-space Virtiofs file-server daemon, \texttt{virtiofsd}~\cite{virtiofsd}. On the client side, DPC adds roughly 3.3,KLoC to the kernel, and the cache directory contributes about 6,KLoC, measured with \texttt{cloc}~\cite{adanial_cloc}. All interactions with the page cache occur through existing file-system and address-space operation hooks, and the cache directory is exposed to clients via a small set of DPC-specific FUSE operations.



The Page Directory entry in \cref{fig:directory} is compact. For a 32-node cluster, each entry includes an 8\,b status word encoding the protocol state (3\,b) and node ID (5\,b) of the current owner, plus 52\,b each for the page's file offset and the owner's page-frame number (PFN), for a total of 14\,B per entry.

We next detail different aspects of the DPC client implementation and its interaction with the cache directory.


\begin{table}[t]
\captionsetup{aboveskip=4pt, belowskip=0pt}
  \scriptsize
  \centering
    \caption{DPC FUSE operations for the client and directory.}
  \begin{tabular}{ll}
    \toprule
    Opcode & Purpose \\
    \midrule
    \texttt{FUSE\_DPC\_READ} & Read + directory lookup (miss handling). \\
    \texttt{FUSE\_DPC\_LOOKUP\_LOCK}  & Batched \textsf{WR\_PREP\_LOCK} over a write range. \\
    \texttt{FUSE\_DPC\_UNLOCK}  & Commit pages (\textsf{E} $\rightarrow$ \textsf{O}) and publish PFNs. \\
    \texttt{FUSE\_DPC\_BATCH\_INV} & Owner-initiated batched invalidation (local reclaim). \\
    \texttt{FUSE\_DIR\_INV} & Directory initiated batched invalidation  request. \\
    \texttt{FUSE\_DPC\_INV\_ACK} & High-priority ACKs for directory invalidation. \\
    \bottomrule
  \end{tabular}

  \label{tab:dpc-fuse}
\end{table}


\subsection{DPC discovery}

During boot and mount, the Virtiofs driver performs a feature negotiation with the storage server. If the virtio device advertises DPC support, it is issued a unique DPC node identifier and descriptors or remote memory regions exported by
all other DPC nodes (physical address ranges). For mounts without DPC support, the FS Shim falls back to unmodified Virtiofs behavior and all DPC code remains dormant.

\subsection{DPC-specific FUSE operations}
DPC extends Virtiofs with a small set of DPC-specific FUSE opcodes (\cref{tab:dpc-fuse}) that drive the directory state machine. Each opcode carries a batch of fixed-size 64\,B page descriptors, so many pages are handled per round trip.

\mypar{Read path}
DPC reuses Linux's buffered read-miss handling. On a page-cache miss for a DPC-backed file, the VFS allocates a page, inserts it into the page cache, and calls Virtiofs. The DPC MM batches such misses and issues a single \texttt{FUSE\_DPC\_READ} with inodes, offsets, and PFNs. For each page, the DPC directory consults the state machine (\cref{fig:fsm}). If all nodes are in \textsf{I}, it grants \textsf{E} to the requester, arranges for the backing store to DMA into the provided PFN, then promotes the page to \textsf{O} on that node. If another node is already in \textsf{O}, the directory moves the requester to \textsf{S}, and returns the owner's node ID and PFN; the client drops the preallocated page and inserts the remote frame into the page cache. Preallocation is needed only for pages that the requester becomes a new owner---to provide a DMA target without another round trip. For remote hits, discarding the unused page is cheap relative to the avoided disk access.

\mypar{Write path}
The generic Linux buffered-write path iterates over the target range page by page, looking up each page in the page cache and allocating a new page if needed before copying user data. For DPC mounts, we keep this iteration but change how cache misses are prepared and committed. To ensure strong coherence, write preparation and commit are decoupled into two steps and batched over contiguous runs of missing pages.  The kernel first identifies which pages in the write range are already present locally. For the remaining pages, it issues a batched \texttt{FUSE\_DPC\_LOOKUP\_LOCK}. For each page in the batch, the directory checks the page state, and transitions it to \textsf{S} or \textsf{E} according to whether is it remotely available or invalid (\cref{subsec:directory}). A page in \textsf{E} is committed to the \textsf{O} state when the node issues a \texttt{FUSE\_DPC\_UNLOCK}.

\subsection{Page reclamation and invalidation}

DPC integrates with Linux’s memory reclamation logic without modifying it, using file-system and address-space callbacks. When the local kernel decides to reclaim a file-backed page, the reclamation path must first invalidate any remote mappings so that no other node continues to access the page after it is freed.

To allow invalidation notifications to the clients, DPC extends the in-kernel Virtiofs implementation with dedicated notification and acknowledgement virtqueues for DPC, so that directory-initiated invalidations and
their acknowledgments are carried on separate paths from regular I/O.

\mypar{Locally initiated reclamation}
Linux's reclaim path walks LRU lists of candidate pages and, for file-backed pages, invokes the file system's \texttt{release\_folio} callback before removing a page from the page cache.
Only pages backed by local DRAM are candidates for reclaim, since dropping a remote mapping does not free local memory.
When the reclaim path selects a page that belongs to a DPC mount, the traditional kernel handler unmaps it from all process page tables, and DPC subsequently
 enqueues it on a per-CPU invalidation batch list. When a batch reaches a threshold (e.g., 32 pages), DPC issues a single \texttt{FUSE\_DPC\_BATCH\_INV} request and
instructs the kernel to retain them on the LRU lists until the directory completes invalidation. This is similar to how writeback occurs for dirty pages.

The directory processes \texttt{FUSE\_DPC\_BATCH\_INV} by sending invalidation requests (\texttt{FUSE\_DIR\_INV}) to the notification queues of all nodes that map the page (\textsf{S}).
Those nodes tear down their mappings and send a notification \texttt{FUSE\_DPC\_INV\_ACK} to the high-priority virtqueue, including a dirty bit that indicates if they observed the page as dirty locally. 
This mirrors intra-node behavior, where multiple page table entries may mark a single physical page dirty but writeback happens only once.
Once the directory receives all ACKs, it replies to the original invalidation request.


Upon receiving this reply, DPC adjusts its accounting for each invalidated page, writes back the page if it is dirty, and moves it to the tail of the inactive LRU list. On the next pass of the kernel's reclaim, these pages are reclaimed first, similar to newly cleaned pages. I/O requests to a page undergoing reclamation are temporarily blocked and retried once reclamation completes, ensuring that no request is served from a page that is being freed.

\mypar{Remotely initiated invalidation}
In the case that an owner node \textsf{A} wants to reclaim a local page that  node \textsf{B} remotely maps, the cache directory sends a \texttt{FUSE\_DIR\_INV} notification to \textsf{B}'s notification virtqueue. 
The Notification Manager on \textsf{B} then locates the remote page corresponding to each (inode, offset) in the notification, and asks DPC MM to unmap the page from all process page tables and remove it from the page cache.
It then sends a \texttt{FUSE\_DPC\_INV\_ACK} that contains the dirty bit and page descriptors in the dedicated acknowledgment virtqueue. 
Using the normal request virtqueue for invalidation acknowledgments could risk deadlock under load, as invalidation handlers during multi-node concurrent invalidations may be blocked waiting for acknowledgments that are themselves 
queued in the same worker pool.

\subsection{Remote MM}
\label{subsec:remote-mm}

DPC relies on CXL to expose a portion of each node’s DRAM as fabric-addressable memory that other nodes can access with coherent loads and stores. In Linux, we model each peer’s exported DRAM range as device memory in \texttt{ZONE\_DEVICE}, managed by a \texttt{dev\_dax}-style~\cite{dev_dax} driver we implement, called \texttt{dpc\_dax}. This driver registers these ranges as device pages so 
that the kernel can obtain \texttt{struct page} descriptors and PFNs for remote frames, while preventing the OS from treating those frames as regular allocatable memory.




During initialization, a DPC Client learns, for each remote node, the physical base address and size of the DRAM range it exports over CXL. \texttt{dpc\_dax} reserves the corresponding PFN ranges, initializes a \texttt{dev\_pagemap} for each, and invokes \texttt{memremap\_pages()} to populate \texttt{ZONE\_DEVICE} with one \texttt{struct page} per 4\,KB frame. When the directory instructs a client to map a remote page identified by a node ID and PFN, \texttt{dpc\_dax} resolves this identifier to the local PFN in the appropriate \texttt{ZONE\_DEVICE} range, and the resulting page is inserted into the page cache and handled like a local page. 

\section{Discussion}
\label{sec:discussion}

\mypar{Security} Exposing host DRAM over CXL must not allow arbitrary remote access. When host DRAM backs the DPC cache, only the page cache frames that are currently enrolled in DPC should be reachable from the CXL fabric. A practical realization is to have the host-side CXL adapter expose a single contiguous CXL.mem region that covers the superset of potential DPC frames, and internally augments it with a page-granular permission bitmap. Functionally, this behaves like a simple identity-mapping IOMMU that implements a fast per-page access check without translation and can be integrated into the CXL card that is required in the host in any case. Mapping a page into DPC sets its bit; unmapping clears it; all other host frames remain private, so the fabric can only reach page cache frames that DPC explicitly exports.

\mypar{Liveness} DPC's reclamation protocol requires invalidating existing remote mappings before a DPC-backed page can be reclaimed. The reclaiming node sends an invalidation request to the cache directory, which forwards it to all nodes that cache the page and waits for acknowledgments. A failed or misbehaving node that never responds would block eviction.

To avoid this, the directory tracks client liveness: when a node becomes unresponsive, the directory marks it failed, stops waiting for its acknowledgments, removes it from all sharer sets, and completes any pending invalidations so eviction can finish. In parallel, the fabric manager unregisters the node from the cluster, so it can no longer issue \texttt{CXL.mem} requests to DPC-exported frames. Symmetrically, if a client does not receive responses from the cache directory within a configured timeout, it treats the directory as failed, disconnects from DPC, invalidates its own remote mappings, and reclaims its pages using the normal local page cache policy.

\mypar{Fault tolerance and durability} DPC treats CXL-backed DRAM strictly as a page cache. A node failure therefore primarily affects cached contents. Loss of clean pages simply shrinks effective cache capacity until the working set is repopulated. Loss of dirty pages has the same semantics as in a conventional write-back page cache: updates that were resident only in DRAM and not yet flushed can be lost.

DPC's design admits deployment policies that constrain where dirty data may reside. One conservative configuration keeps writable pages local: once a node dirties a page, the directory can stop advertising that page as a remotely mappable cache entry. Remote nodes then either read a clean copy or fetch data directly from storage, so a remote failure can only discard read-only cache state. Other  configurations may allow dirty pages to be shared via DPC to maximize performance, however, a failure can lose a subset of unflushed updates, similar to other distributed systems with write-back caching.

\mypar{Relaxed consistency DPC} DPC also implements a \emph{relaxed coherence mode}, that deliberately trades strong consistency for performance on buffered I/O. In this mode, each node may keep its own writable copy of a page in the local page cache. The directory tracks nodes caching pages that have entered DPC, but pages created purely by local buffered writes without prior DPC access remain local-only and are not tracked. Dirty data is reconciled only when pages are written back to the backing storage, yielding weak semantics similar to NFS and other remote, buffered-I/O file systems.
\section{Evaluation}
\label{sec:evaluation}

Our evaluation demonstrates the latency, bandwidth, and IOPS benefits of DPC over different remote file systems, as well as end-to-end application performance. We also quantify the overheads introduced by DPC's coherence and metadata operations.  To evaluate DPC we deploy it on a CXL emulated multi-host
environment running on a commodity ARM server.

Because production CXL~3.0  hardware is not yet available,
our evaluation platform cannot rely on real CXL fabrics.
Instead, we emulate CXL in QEMU~\cite{qemu} and use NUMA nodes with the hardware cache-coherence protocol
of the underlying system to provide a realistic model of coherent, byte-addressable shared memory, as performed by prior works~\cite{cxlnuma1,li2023pond,liu2024dissecting,Sun2023demystifyingcxl}. 

\mypar{Testbed}
We evaluate DPC on a dual-socket 64-bit ARMv8.2 server (model not specified for anonymization).
Each socket integrates two compute dies, with each incorporating 32 cores at 2.6\,GHz, for a total of 128 physical cores.
The system has 256\,GiB of DDR4-2933 memory, evenly distributed across the four NUMA nodes
that correspond to the compute dies.
Persistent benchmark data resides on a 2\,TB RAID-0 array built from 4$\times$480\,GB SAS~3.0 enterprise SSDs
(each SSD provides 1\,GB/s sequential read and 1\,GB/s sequential write bandwidth).
The OS image and system logs reside on a separate SSD to avoid interference with the data path.

\subsection{Experimental methodology}

\begin{figure}
\captionsetup{aboveskip=4pt, belowskip=0pt}
\vspace{-1.0\baselineskip} 
    \centering
    \includegraphics[width=\linewidth]{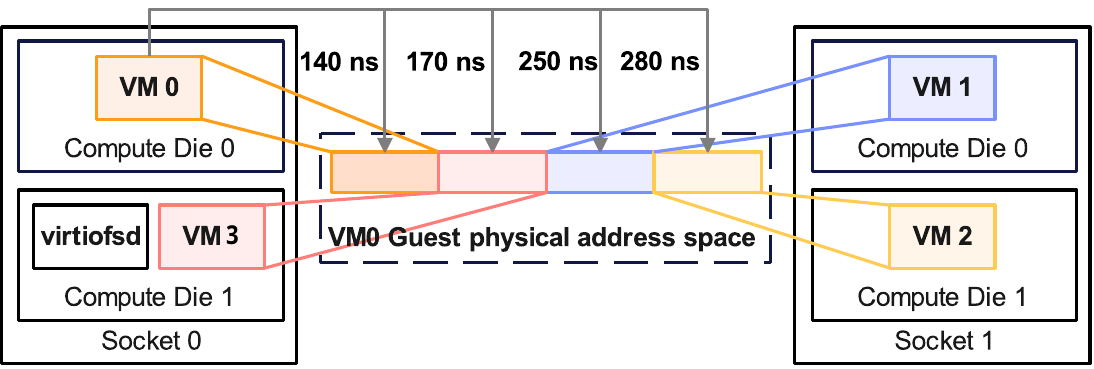}
    \caption{Experimental setup for CXL emulation, with memory access latencies.}
    \label{fig:experimental_setup}
\end{figure}
We emulate a CXL memory fabric on a single host using QEMU (v8.2) and KVM~\cite{kvm}.
Each compute node is realized as a separate QEMU virtual machine (VM).
Both the hypervisor and guests run Ubuntu~22.04 with Linux kernel~6.12.5. 
Guest kernels include our full DPC implementation.

\begin{figure*}[h!]
\captionsetup{aboveskip=4pt, belowskip=0pt}
\centering
\subfloat[a][Latency \texttt{bs=4k,qd=1,jobs=1}]
{\includegraphics[trim={0pt 8pt 0pt 11pt},clip,width=0.322\textwidth]{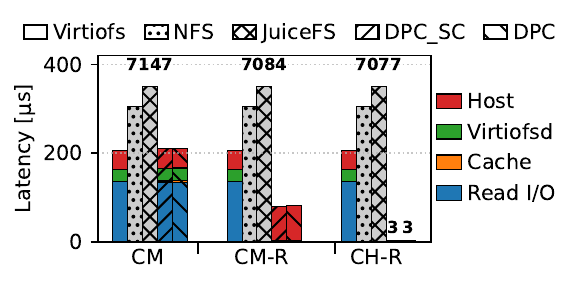} \label{fig:read-aio-lat}}
\subfloat[b][Bandwidth \texttt{bs=128k,qd=32,jobs=8}]{
 \includegraphics[trim={0pt 9pt 0pt 11pt},clip,width=0.322\textwidth]{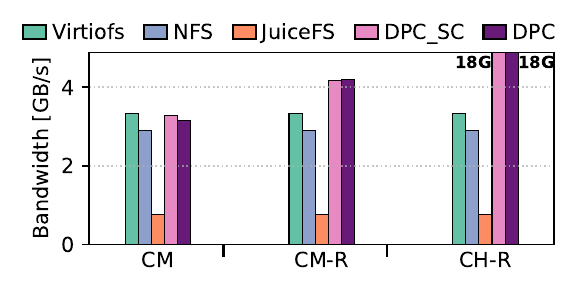}
    \label{fig:read-aio-bw}
}
\subfloat[c][IOPS \texttt{bs=4k,qd=32,jobs=8}]{
\includegraphics[trim={0pt 9pt 0pt 11pt},clip,width=0.322\textwidth]{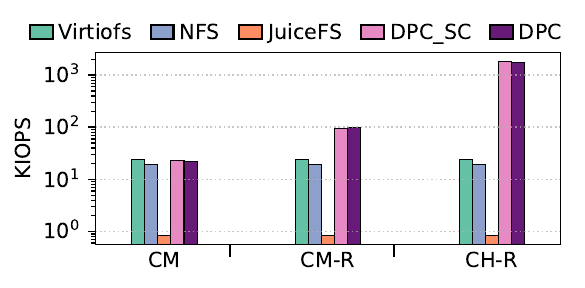}
    \label{fig:read-aio-iops}
}
\caption{Read latency, bandwidth, and IOPS using \texttt{fio} with \texttt{libaio}.}
\end{figure*}

\begin{figure*}[h!]
\captionsetup{aboveskip=4pt, belowskip=0pt}
\vspace{-1.0\baselineskip} 
\centering
\subfloat[a][Latency \texttt{bs=4k,qd=1,jobs=1}]
{\includegraphics[trim={0pt 8pt 0pt 11pt},clip,width=0.322\textwidth]{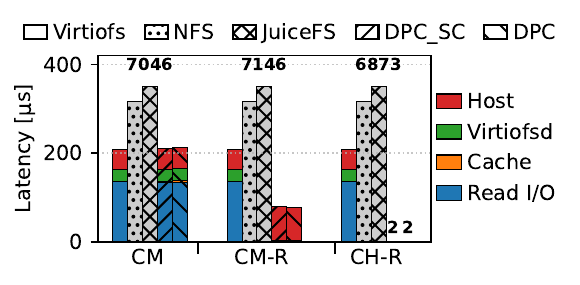} \label{fig:read-map-lat}}
\subfloat[b][Bandwidth \texttt{bs=128k,qd=32,jobs=8}]{
 \includegraphics[trim={0pt 9pt 0pt 11pt},clip,width=0.322\textwidth]{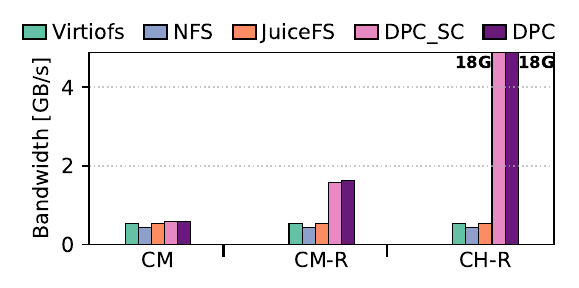}
    \label{fig:read-map-bw}
}
\subfloat[c][IOPS \texttt{bs=4k,qd=32,jobs=8}]{
\includegraphics[trim={0pt 9pt 0pt 11pt},clip,width=0.322\textwidth]{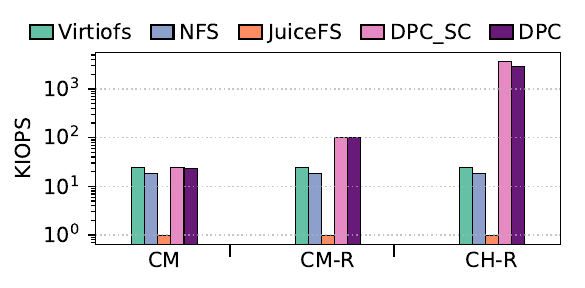}
    \label{fig:read-map-iops}
}
\caption{Read latency, bandwidth, and IOPS using \texttt{fio} with \texttt{mmap}.}
\label{fig:read-map}
\end{figure*}

Each VM is configured with 16~vCPUs and 32\,GB of RAM. 
vCPUs of a VM are pinned to distinct physical cores within a single compute die,
and the VM's memory is allocated from the NUMA node local to that die.
Different compute VMs are pinned to different dies, yielding up to four isolated emulated hosts.
We pre-fault and pin all guest memory at boot time to allocate all the VMs' memory and
to populate all second-level address translation tables.

The storage server is implemented by \texttt{virtiofsd}~\cite{virtiofsd} (v1.13.1) on the host, and maps an XFS file system 
that is mounted on the RAID-0 LVM array to the  VMs via Virtiofs.
The \texttt{virtiofsd} daemon uses a thread pool of size~2. 

\cref{fig:experimental_setup} illustrates how we emulate CXL~3.0-like memory sharing between hosts.
Each VMs' memory is backed by a distinct memory-backend file~\cite{mem-backend-file} on the host that is loaded entirely into memory.
For every VM, we attach the memory-backend files of the other VMs as CXL devices behind CXL root ports in QEMU, where they are handeled by
 the \texttt{dpc\_dax} driver.
As all CXL-backed regions ultimately reside in shared host DRAM, 
loads and stores from different VMs observe a coherent view of these regions,
mirroring the coherence guarantees of a real CXL deployment.
This setup lets us evaluate DPC and remote page access paths under a realistic CXL memory-sharing model
while remaining independent of any specific vendor's early CXL hardware.

\mypar{Configurations} 
We evaluate DPC against the following system configurations, all backed by the same RAID-0:

\myparij{(a)~Virtiofs:} Baseline configuration with unmodified Virtiofs. 

\myparij{(b)~JuiceFS:} JuiceFS~\cite{juicefs_juicedata_2025} (version 1.3), a distributed POSIX file system accessed via its user-space client.
The compute VMs mount JuiceFS over virtio-net~\cite{virtio_spec_1_3_2023}  interface with its local memory
write-back cache enabled.

\myparij{(c)~NFS:} NFSv4.1~\cite{nfs} with the Linux client.
The compute VMs mount an NFS export of the same XFS file system.

\myparij{(d)~DPC\_SC:} Base DPC with strong consistency (\cref{sec:design}).

\myparij{(e)~DPC:} DPC with relaxed consistency,
analogous to an NFS-style weak consistency model~\cite{nfs} as described in~\cref{sec:discussion}.

\subsection{Microbenchmarks} 
All microbenchmarks use \texttt{fio}~\cite{fio_axboe_2022} (v3.28) and run inside the VMs.
\texttt{fio} is executed with \texttt{direct=0} so that all operations go through the page cache. We measure three metrics: I/O latency, bandwidth, and IOPS.
For latency, \texttt{fio} issues random 4\,KB I/O operations with a single job and a queue depth of 1
for 30\,s. To measure bandwidth, 8 jobs perform I/O on distinct files sequentially
using 128\,KB blocks and queue depth of 32, and measure the aggregate throughput across all jobs. For IOPS, we use 8 jobs, each randomly accessing 4\,KB blocks in a private per job 1\,GB file with a queue depth of 32.

\begin{figure*}[h!]
\captionsetup{aboveskip=4pt, belowskip=0pt}
\centering
\subfloat[a][Latency \texttt{bs=4k,qd=1,jobs=1}]
{\includegraphics[trim={0pt 8pt 0pt 11pt},clip,width=0.322\textwidth]{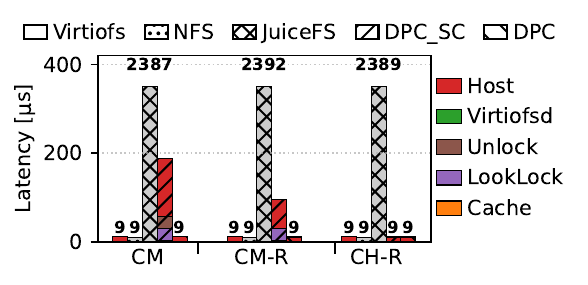} \label{fig:write-aio-lat}}
\subfloat[b][Bandwidth \texttt{bs=128k,qd=32,jobs=8}]{
 \includegraphics[trim={0pt 9pt 0pt 11pt},clip,width=0.322\textwidth]{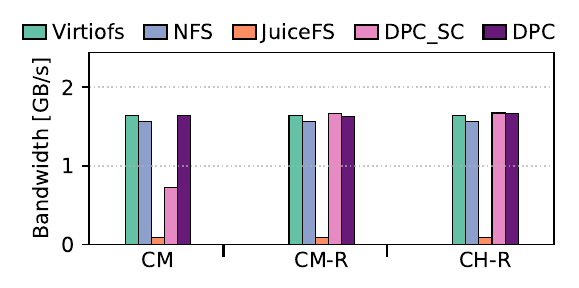}
    \label{fig:write-aio-bw}
}
\subfloat[c][IOPS \texttt{bs=4k,qd=32,jobs=8}]{
\includegraphics[trim={0pt 9pt 0pt 11pt},clip,width=0.322\textwidth]{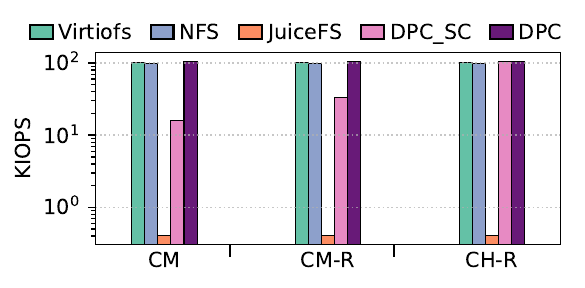}
    \label{fig:write-aio-iops}
}
\caption{Write latency, bandwidth, and IOPS using \texttt{fio} with \texttt{libaio}.}
\label{fig:write-aio}
\end{figure*}

\begin{figure*}[h!]
\captionsetup{aboveskip=4pt, belowskip=0pt}
\vspace{-1.0\baselineskip} 
\centering
\subfloat[a][Latency \texttt{bs=4k,qd=1,jobs=1}]
{\includegraphics[trim={0pt 8pt 0pt 11pt},clip,width=0.322\textwidth]{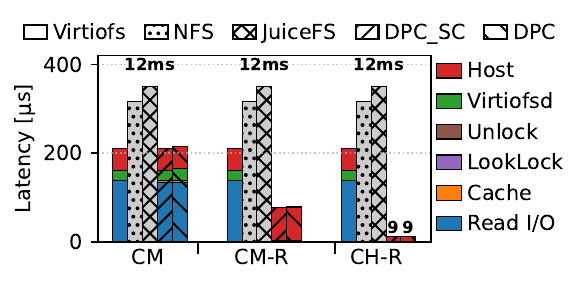} \label{fig:write-map-lat}}
\subfloat[b][Bandwidth \texttt{bs=128k,qd=32,jobs=8}]{
 \includegraphics[trim={0pt 9pt 0pt 11pt},clip,width=0.322\textwidth]{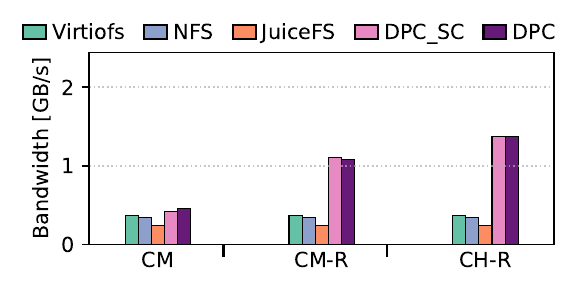}
    \label{fig:write-map-bw}
}
\subfloat[c][IOPS \texttt{bs=4k,qd=32,jobs=8}]{
\includegraphics[trim={0pt 9pt 0pt 11pt},clip,width=0.322\textwidth]{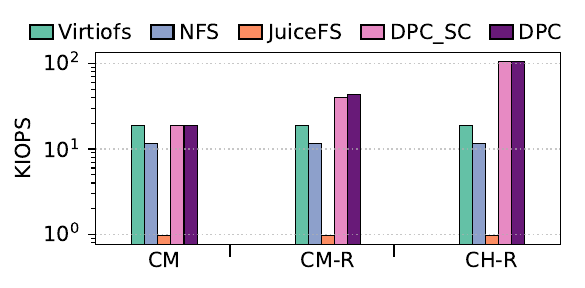}
    \label{fig:write-map-iops}
}
\caption{Write latency, bandwidth, and IOPS using \texttt{fio} with \texttt{mmap}.}
\label{fig:write-map}
\end{figure*}

We evaluate three cache-residency scenarios:\\
\myparij{CM (Cache Miss).} The target page is not cached on any node, so the accessing node must fetch data from the storage server.

\myparij{CM-R (Cache Miss-Remote).} The target page is not cached locally, but it is cached on a remote node in the DPC cluster.
\myparij{CH-R (Cache Hit-Remote).} The benchmark node has previously mapped the page remotely via DPC.

For CM-R, we warm the cache on a remote node (VM~0 in \cref{fig:experimental_setup}) and run the benchmark on another (VM~2); the first access installs a remote mapping to the owner’s frame. For CH-R, we reuse the CM-R setup but we ensure that the benchmark node has remote mappings so accesses hit the remote DRAM directly. We run each experiment with two \texttt{fio} engines. The \texttt{libaio} engine issues buffered asynchronous I/O via \texttt{read}/\texttt{write}. The \texttt{mmap} engine memory-maps the file and issues loads/stores. Evaluating both engines allows us to contrast DPC's benefits for syscall based I/O and memory-mapped access.

%

\subsubsection{Read microbenchmarks with \texttt{libaio}}
\label{eval:seq_read}


\cref{fig:read-aio-lat} shows the latency breakdown, bandwidth, and IOPS for buffered reads with the \texttt{libaio} engine across the CM, CM-R, and CH-R scenarios. Under CM, all systems fetch data from the storage server; for Virtiofs, the mean latency is 205,\textmu s. DPC and DPC\_SC add no measurable overhead, as directory communication piggybacks on the read and hash-table lookups are negligible relative to media latency. In the CM-R scenario, DPC and DPC\_SC leverage the cached pages residing in other nodes.
The first access performs a directory lookup, installs a remote mapping in the page cache, and then reads the data from remote DRAM instead of the SSD array. This reduces latency by 2.6$\times$ compared to Virtiofs.
The ``Host'' component in the breakdown (compute node kernel time) slightly increases, reflecting the extra work to replace the pre-allocated
page cache entry with a remote PFN. As described in \cref{sec:implementation}, pre-allocated pages are required  to serve as targets for read operations in cases the data is not available on a remote node. Once the remote mapping is established, CH-R corresponds to subsequent reads that hit directly in remote DRAM. For DPC and DPC\_SC, the latency is dominated by the syscall overhead, \texttt{fio}, and the remote memory access. 

\cref{fig:read-aio-bw} shows the corresponding bandwidth results.
DPC and DPC\_SC achieve up to 1.3\,$\times$ higher read bandwidth than Virtiofs in CM-R 
by serving large sequential reads from remote DRAM rather than the SSD array. DPC's gains in CM-R are constrained
by the cost of replacing pre-allocated page cache pages,
 as it requires
performing these operations at a 4\,KB granularity.
At CH-R, DPC and DPC\_SC achieve a 4.5$\times$ speedup over Virtiofs. The IOPS results, shown in~\cref{fig:read-aio-iops} follow the same trend: DPC variants substantially increase the 
IOPS in CM-R and CH-R against Virtiofs, up to 72.8$\times$.

\subsubsection{Read microbenchmarks with \texttt{mmap}} 
\cref{fig:read-map} presents the same experiments using \texttt{fio}'s \texttt{mmap} engine.
Each job memory maps its file and issues loads that trigger page faults on 4\,KB boundaries.
Linux readahead prefetches additional pages, but less than the benchmark's explicit 128\,KB read calls in the
\texttt{libaio} experiments. As a result, both DPC and the baselines perform smaller granularity transfers from storage
in the bandwidth test scenario.

Overall, the \texttt{mmap} results for read follow the same trends as the \texttt{libaio} experiments: 
in CM-R and CH-R, DPC and DPC\_SC  outperform Virtiofs, NFS, and JuiceFS by serving requests
from remote DRAM instead of the SSD array, and their latency is dominated by remote-memory access and
page-fault handling rather than storage.
Across both engines, JuiceFS exhibits noticeably higher latency and lower bandwidth than Virtiofs and NFS,
consistent with prior findings~\cite{juicefs_report}.

\subsubsection{Write microbenchmarks with \texttt{libaio}}
\cref{fig:write-aio-lat} shows the latency breakdown for buffered writes using
\texttt{libaio}. Virtiofs, NFS, and DPC achieve low write latencies across all scenarios: with a buffered 4\,KB write, the kernel allocates a page in the page cache and copies the data without communication with the storage server. The measured latency therefore reflects page allocation and the in-memory copy. DPC\_SC behaves differently as it must enforce strong coherence. Before completing a write, DPC\_SC contacts
the directory to determine whether another node holds a copy of the target page and, if necessary, 
participates in the two-step protocol (\cref{subsec:directory}) to acquire exclusive ownership.
This additional control-plane work increases latency in CM, with a write latency of 195us.
In CM\_R, this cost is partially offset:
when another node already owns the page, DPC\_SC can avoid allocating a new local page and instead map
the remote page directly. Therefore, DPC\_SC can  skip the second step of the two-step ownership,
as the page is not locally owned.

The bandwidth and IOPS in \cref{fig:write-aio-bw,fig:write-aio-iops} are 
limited by the CPU in the \texttt{write} syscall rather than by the storage backend. All configurations 
converge to similar bandwidths, but DPC\_SC pays an IOPS penalty in CM and CM-R due to per-page strong-coherence checks, which are unavoidable. 
In CM-R, DPC\_SC is able to eliminate the bandwidth gap by batching directory operations over 128\,KB extents, which allows it to hide much
of the directory latency behind the data copy when issuing larger writes.

\subsubsection{Write microbenchmarks with \texttt{mmap}} 
\cref{fig:write-map} evaluates writes using \texttt{fio}'s \texttt{mmap} engine. Here, each store to a mapped region
targets a 4\,KB page. If the page is not resident, the kernel must first fault it in (a read) before writing the data. Consequently, write latencies in CM 
are higher in \texttt{mmap} than the \texttt{libaio} engine, and resemble the read latencies from the previous experiments. In CM-R, DPC and DPC\_SC again benefit from remote caching: they can satisfy the initial page faults from remote DRAM
instead of disk, then apply writes directly, whereas Virtiofs, NFS, and JuiceFS continue to fetch data from the storage
server. The bandwidth and IOPS patterns mirror those of the read \texttt{mmap} experiments but incur additional overhead from
copying data into the page cache on writes. In CM-R, DPC and DPC\_SC achieve 2.7$\times$ lower latency, 2.8$\times$ higher IOPS, and 3$\times$ higher bandwidth against Virtiofs. In CH-R, DPC and DPC\_SC achieve 23.3$\times$ lower latency, 18.3$\times$ higher IOPS and 3.7$\times$ higher bandwidth against Virtiofs.

Taken together, these microbenchmarks show that DPC preserves baseline behavior when no cached copies are available (CM), and DPC\_SC pays a per-write overhead for stronger
consistency guarantees.
DPC and DPC\_SC substantially improving performance whenever data can be reused from remote page caches (CM-R and CH-R). 

\begin{figure*}[t]
\captionsetup{aboveskip=4pt, belowskip=0pt}
  \centering
  \vspace{10pt}
  \includegraphics[width=1\textwidth]{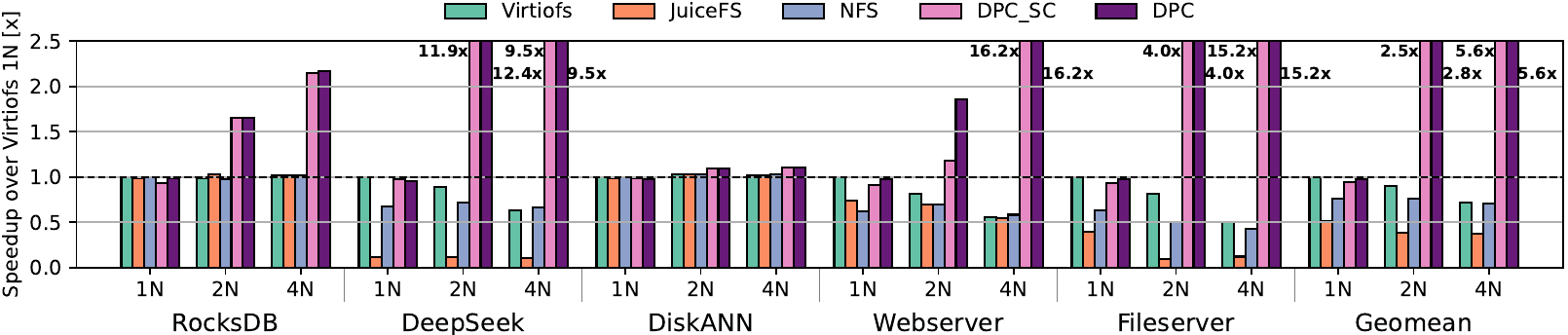}
  \caption{Relative speedup of application benchmarks over single-node Virtiofs. For multi-node configurations, bars report per-node performance averaged across all VMs (higher is better).}
  \label{fig:combined_speedup}
\end{figure*}

\subsubsection{Page reclamation overheads}
\label{sec:thrash}
When the local kernel reclaims a DPC-backed page cache entry under memory pressure, the DPC cache directory must first ensure that no other node still maps that page. As described in \cref{sec:design}, the reclaim path sends an invalidation request to the directory, which coordinates any required remote unmapping before the page is freed locally. This coordination is inherent to a distributed page cache and does not occur in a baseline Virtiofs configuration. To quantify this coordination cost in isolation, we measure the latency of a synchronous single-page invalidation under Virtiofs and under the DPC configurations. In the DPC experiment, the page is also mapped in other VMs. In the Virtiofs baseline, invalidation completes in 11\,us and involves only local operations, whereas in DPC it takes 99.7\,us as the directory must be consulted and remote sharers must tear down their mappings before reclaim.

In practice, DPC hides this higher per-page latency behind the asynchronous, batched mechanism described in \cref{sec:implementation}. To evaluate the end-to-end effect under memory pressure, a VM is configured with 2\,GB of memory and re-runs the sequential read bandwidth experiment from \cref{eval:seq_read} on large files, thereby thrashing system memory and triggering frequent page reclamation. Across Virtiofs and DPC variants, the measured bandwidth remains unchanged: sequential reads are limited by the storage and reclamation paths, and DPC overlaps asynchronous invalidations with page scanning and reclamation. Thus, although synchronous invalidation is more expensive under DPC than under Virtiofs, the batched design prevents invalidation from becoming a bottleneck during page-frame reclamation, as we also show in~\cref{sec:application}.

\subsection{Application benchmarks}
\label{sec:application}
All setups are evaluated with the following applications:

\myparij{RocksDB}~\cite{rocksdb} is a key-value store. The benchmark uses a 60\,GB database with 24-byte keys and 1\,KB values, and runs \texttt{db\_bench} \cite{rocksdb} in \texttt{readrandom} for 300\,s.\\ \emph{Working-set size:} 60\,GB. \emph{Metric:} queries per second (QPS).

\myparij{DeepSeek} denotes CPU-based inference using the DeepSeek-V2-Lite-Chat model~\cite{deepseekai2024deepseekv2strongeconomicalefficient}. The model weights reside in a single 30\,GB file that is memory-mapped into the process, and each run uses a fixed chat-style prompt and random seed to produce a deterministic token sequence.\\
\emph{Working-set size:} 30\,GB. \emph{Metric:} tokens per second (TPS).

\myparij{DiskANN}~\cite{diskann} is a disk-based approximate nearest-neighbor index. The benchmark uses a 40\,GB OPQ-quantized index over the SIFT100M~\cite{NNPQ} dataset (100\,M float32 vectors, 128 dimensions) and evaluates 1\,M L2-distance queries, each returning 100 neighbors.\\ \emph{Working-set size:} 40\,GB. \emph{Metric:} queries per second (QPS).

\myparij{Webserver} is based on Filebench's \texttt{webserver}~\cite{filebench}, which models a static-content server that repeatedly opens, reads, and closes small files, and appends to log file. The fileset consists of 2000 files (each <16\,MiB) scaled to a 32\,GB logical dataset and is run for 10 minutes.\\ \emph{Working-set size:} 32\,GB. \emph{Metric:} geometric mean throughput of all operations.

\myparij{Fileserver} uses Filebench's \texttt{fileserver} configuration, emulating multiple users creating, reading, writing, appending, and deleting files in a shared directory tree over a 2000 file set (sizes from KBs up to 32\,MB) for 10 minutes.\\ \emph{Working-set size (approx.):} 32\,GB. \emph{Metric:} geometric mean throughput of all operations.

To evaluate multiple machines using DPC, we vary the number of nodes that concurrently run the same application over a shared dataset. This configuration models deployments where several front-end instances of the same service access shared state and stresses DPC's control plane: the cache directory, lookup path, and cross-node invalidation handling while other VMs are also performing similar operations.

\cref{fig:combined_speedup} summarizes the application-level performance normalized to a single Virtiofs node as the number of nodes increases. For multi-VM configurations, the per-node metric is computed as the average across all VMs. Across all workloads, per-node performance with DPC and DPC\_SC does not decrease when adding more nodes. This behavior indicates that DPC’s directory operations and cross-node coordination do not introduce a scalability bottleneck in these scenarios.

With a single node, both DPC variants match Virtiofs within 2\% across all workloads, even though each working set exceeds the page cache capacity of a single machine and triggers frequent reclamation. This result shows that DPC's directory lookups and invalidation handling during reclamation do not degrade performance, even under sustained page cache thrashing, as also evaluated in~\Cref{sec:thrash}.

DeepSeek, Webserver, and Fileserver have relatively low computational intensity against the other workloads. In these workloads, DPC's ability to coordinate page reuse across nodes yields the most pronounced gains. It further improves application performance by reducing memory pressure due to the distributed page cache. DiskANN and RocksDB spend more CPU time per I/O, so I/O is a smaller fraction of runtime.

When the combined page caches of 2 or 4 nodes are sufficient to hold the full working set, the benefits of DPC become more apparent. In these configurations, DPC and DPC\_SC substantially outperform Virtiofs, achieving up to 12.4$\times$, 16.2$\times$, and 15.2$\times$ higher throughput for DeepSeek, Webserver, and Fileserver, respectively. These gains arise because once one node faults a page into memory, other nodes can reuse it via DPC rather than refetching it from storage, effectively turning aggregate DRAM across the DPC cluster into a shared cache for application data.

Across several applications, DPC\_SC often trails DPC slightly because write operations incur extra coordination, but the impact is modest: in the 2-node configuration, DPC attains a 2.8$\times$ geomean speedup over Virtiofs across applications, compared to 2.5$\times$ for DPC\_SC.

\section{Related Work}

\textbf{\small Memory pooling and sharing via CXL.} 
Prior work builds CXL-based memory pools to extend the memory available to a host. DirectCXL~\cite{atc22-directcxl} prototypes a CXL memory pool atop an FPGA. TPP~\cite{asplos23-tpp} proposes an OS-level, application-transparent page-placement mechanism for CXL-enabled tiered memory, and Pond~\cite{li2023pond} orchestrates page allocation across local and pooled memory. In these designs, each memory page is still owned and accessed by a single host, whereas DPC exposes a shared page cache across hosts. DPC is complementary to these mechanisms and can build on them: migrate hot pages or maintain read-only replicas to mitigate cross-host hotspots.

Furthermore, several works study memory sharing via CXL, where a memory page is accessible to multiple hosts. CXL-SHM~\cite{sosp23-cxlshm} automatically manages objects in shared CXL memory, and TrEnv~\cite{sosp23-cxlshm} rapidly restores serverless function state from it. The database community explores building a shared buffer pool on CXL memory~\cite{icde24-cxl-db, sigmod25-cxl-database}, while FamFS~\cite{hipcw24-famfs} instances on separate hosts construct a file system atop shared CXL memory rather than the general page cache. In contrast, DPC specifically targets the page cache.

\textbf{\small Page cache optimization.}
Prior work optimizes the page cache along two main dimensions:\\
\textit{Scalability.} Zheng  \etal{}  propose a user-space, set-associative page cache~\cite{sc13-parallel-page-cache}, while others reduce contention in page-cache data structures to improve scalability~\cite{atc20-fastmap,eurosys24-scalecache}. \\
\textit{Policy.} Some systems integrate online prefetching and stream awareness into the page cache~\cite{atc20-leap,atc24-streamcache}, and others use eBPF to monitor system state and drive paging decisions in kernel or user space~\cite{atc25-pageflex,sosp25-cache-ext}.
DPC explores an aspect different to prior works. It leverages CXL memory pooling to build a distributed page cache shared across multiple hosts.


\section{Conclusion}
We introduced \sysname, an OS-level distributed page cache that treats cluster DRAM as a single cache budget while preserving familiar file-system interfaces and semantics. \sysname combines a page--granular directory with a single-copy invariant and leverages CXL memory semantics so that remote cache hits are served via low-latency, coherent loads and stores rather than software RPCs. Our end-to-end implementation on a CXL-based emulation framework demonstrates that this design improves efficiency: across real-world and representative data-sharing workloads, \sysname outperforms state-of-practice distributed file-access mechanisms by up to 12.4$\times$ (5.6$\times$ geomean). Looking ahead, we see \sysname as a first step toward treating CXL memory fabrics as a general substrate for OS-level, cluster-wide caching and memory management.

\bibliographystyle{plain}
\bibliography{main}

\end{document}